\documentclass[11pt]{article}
\usepackage[a4paper,text={16.8cm,22.4cm}]{geometry}
\usepackage{authblk}
\usepackage{amsmath,amsfonts,slashed,amssymb,tikz,bm,psfrag,graphicx,color,dsfont}
\usepackage{float}
\usepackage[colorlinks=true,linkcolor=red,citecolor=blue]{hyperref}
\usepackage{cite}
\allowdisplaybreaks 
\begin{document}
\begin{titlepage}
\title{Power corrections to the pion transition form factor
from higher-twist  distribution amplitudes of photon }
\author[$a$]{Yue-Long Shen\footnote{shenylmeteor@ouc.edu.cn}}
\author[$b$,$c$]{Jing Gao\footnote{gaojing@ihep.ac.cn}}
\author[$b$,$c$]{Cai-Dian L\"u\footnote{lucd@ihep.ac.cn}}
\author[$a$]{Yan Miao\footnote{782771338@qq.com}}
\affil[$a$]{\it \small College of Information Science and
Engineering, Ocean University of China, Qingdao, Shandong 266100,
P.R. China} \affil[$b$]{\it \small Institute of High Energy
Physics, CAS, P.O. Box 918(4), Beijing 100049,   P.R. China }
 \affil[$c$]{\it \small School of Physics, University of China Academy of Sciences, Beijing 100049,   P.R. China}
\maketitle \vspace{0.2cm}
\begin{abstract}

In this paper we investigate the power suppressed contributions
from two-particle and three-particle twist-4 light-cone
distribution amplitudes (LCDAs) of photon within the framework of
light-cone sum rules.  Compared with leading twist LCDA result, the contribution
from three-particle twist-4 LCDAs is
not suppressed in the expansion by $1/Q^2$, so that the power
corrections considered in this work can give rise to a sizable
contribution, especially at low $Q^2$ region. According to our
result, the power suppressed contributions should be included in
the determination of the Gegenbauer moments of pion LCDAs with the
pion transition form factor.
\end{abstract}
\end{titlepage}

%********************************************************************************
\section{Introduction}
\label{sect:Intro}
%********************************************************************************
As one of the simplest hard exclusive processes, the pion
transition form factor  $F_{\gamma^{\ast} \gamma \to
\pi^{0}}(Q^2)$ at large momentum transfer  is of great importance
in exploring the strong interaction dynamics of hadronic reactions
in the framework of QCD, and to determine the parameters in the
LCDAs of pion. It is defined via the matrix element
\begin{eqnarray}
\langle \pi(p) | j_{\mu}^{\rm em}  | \gamma (p^{\prime}) \rangle =
g_{\rm em}^2 \, \epsilon_{\mu \nu \alpha \beta} \, q^{\alpha} \,
p^{\beta} \, \epsilon^{\nu}(p^{\prime}) F_{\gamma^{\ast} \gamma
\to \pi^0} (Q^2) \,,\, \qquad \epsilon_{0123}= -1
\end{eqnarray}
where $q=p-p^{\prime}$, $p$ and $p'$ refer to the four-momentum of
the pion and the on-shell photon respectively,  the
electro-magnetic current
\begin{eqnarray}
j_{\mu}^{\rm em} = \sum_q \, g_{\rm em} \, Q_q \, \bar q \,
\gamma_{\mu} \, q  \,. \label{em current definition}
\end{eqnarray}In collinear factorization theorem,
pion transition form factor can be factorized into the convolution
of the hard kernel and the leading twist pion LCDA at leading
power of $1/Q^2$
\cite{Lepage:1980fj,Efremov:1979qk,Duncan:1979ny,Rothstein:2003wh},
and the hard kernel has been calculated up to two-loop level
\cite{delAguila:1981nk,Braaten:1982yp,Kadantseva:1985kb,Melic:2002ij}.
At one-loop level,  the factorization formula is written by
\begin{eqnarray}
F_{\gamma^{\ast} \gamma \to \pi^0}^{\rm LP} (Q^2)= \frac{\sqrt{2}
\, (Q_u^2-Q_d^2) \, f_{\pi}}{Q^2} \, \int_0^1 \, d x \, \left[
T^{(0)}_{2}(x)  + T^{(1), \, \Delta}_{2}(x, \mu) \right ] \,
\phi_{\pi}^{\Delta} (x, \mu) + {\cal O}(\alpha_s^2)   \,,
\label{one-loop factorization formula at LP}
\end{eqnarray}
where the  leading twist pion LCDA is defined as
\begin{eqnarray}
\langle  \pi(p) |\bar \xi (y) \, [y, 0] \, \gamma_{\mu} \,
\gamma_5 \, \xi(0) |  0 \rangle = -i \, f_{\pi} \, p_{\mu}  \,
\int_0^1 \, d u \, e^{i\, u \, p \cdot y}  \, \phi_{\pi} (u, \mu)
+ {\cal O}(y^2) \,,
\end{eqnarray}
and the superscript ``$\Delta$" indicates the scheme to deal with
$\gamma_5$ in dimensional regularization which is a subtle problem
in QCD loop diagrams
\cite{Bonneau:1990xu,Collins:1984xc,Larin:1993tq,Martin:1999cc,
Jegerlehner:2000dz,Moch:2015usa,Gutierrez-Reyes:2017glx}.
Employing the trace technique, the $\gamma_5$ ambiguity of
dimensional regularization was resolved by adjusting the way of
manipulating $\gamma_5$ in each diagram to preserve the
axial-vector Ward identity  \cite{Braaten:1982yp}. In a recent
paper \cite{Wang:2017ijn}, the one loop calculation is revisited
by applying the standard OPE technique
\cite{Beneke:2004rc,Beneke:2005gs,Beneke:2005vv} with the
evanescent operator(s) \cite{Dugan:1990df,Herrlich:1994kh}, in
both the NDR and HV schemes for $\gamma_5$ in the $D$-dimensional
space. At one-loop level it has been shown explicitly that the
scheme dependence of the hard kernel and the twist-two pion LCDA
is cancelled out precisely, which guarantees the form factor
$F_{\gamma^{\ast} \gamma \to \pi^0} (Q^2)$ to be free from
$\gamma_5$ ambiguity.

  At leading power
the pion transition form factor has also been studied with
transverse momentum dependent (TMD) factorization approach at
one-loop level \cite{Nandi:2007qx,Musatov:1997pu,Wu:2010zc}, where
the joint resummation of the  large logarithms $\ln^2{{
k_{\perp}^2} /Q^2}$ and $\ln^2 x$  was performed in moment and
impact-parameter space \cite{Li:2013xna}. The prediction of joint
resummation improved TMD factorization approach can accommodate
the anomalous BaBar  measurements \cite{Aubert:2009mc} of
$F_{\gamma^{\ast} \gamma \to \pi^0} (Q^2)$, which have stimulated
intensive theoretical investigations with various phenomenological
approaches as well as lattice QCD simulations (see for instance
\cite{Masjuan:2012wy,Hoferichter:2014vra,Gerardin:2016cqj}). In
Ref. \cite{Radyushkin:2009zg,Polyakov:2009je}, a leading twist
pion LCDA with the non-vanishing end-point behavior was proposed
 to explain the anomalous  BaBar data at high $Q^2$. Later it was found   that
 this method  is
 able to be achieved by introducing  a sizable nonperturbative soft correction
from the TMD pion wave function\cite{Agaev:2010aq}.

To achieve more precise theoretical predictions, power corrections
need to be taken into account especially at low $Q^2$.   In
\cite{Agaev:2010aq,Agaev:2012tm}, the soft correction to the
leading twist contribution is evaluated with the dispersion
approach and found to be crucial to suppress the contributions
from higher Gegenbauer moments of the twist-2 pion LCDAs
\cite{Kroll:2010bf,Li:2013xna}. Furthermore, the subleading power
``hadronic" photon correction can also be taken into account
effectively with dispersion approach. Within this method the
theoretical accuracy for predicting the pion-photon form factor is
improved by including the next-to-next-to-leading order (NNLO) QCD
correction to the twist-2 contribution and the finite-width effect
of the unstable vector mesons in the hadronic dispersion relation
\cite{Bakulev:2002uc,Stefanis:2012yw,Bakulev:2011rp,Bakulev:2012nh,Mikhailov:2016klg}.
Another approach to accommodate the contribution from the
``hadronic photon"  is to introduce the LCDAs of photon. In
\cite{Wang:2017ijn}, the QCD factorization of the correlation
function  for the construction of the LCSRs for the hadronic
photon contribution to the pion-photon form factor is established.
Both the hard matching coefficient and the leading twist photon
LCDAs are  independent of the $\gamma_5$ prescription in
dimensional regularization, and the next-to-leading
logarithmic(NLL) resummation of the  large logarithms was also
perform by solving the renormalization group equations(RGE) in
momentum space. The contribution from the twist-4 pion LCDA is
also calculated at tree level in
\cite{Khodjamirian:1997tk,Wang:2017ijn}. There is strong
cancellation  between this contribution and the contribution from
hadronic structure of photon, which makes the overall power
correction not significant. The LCDAs of photon, including both
two-particle and three-particle Fock state, have been studied up
the twist-4 level \cite{Ball:2002ps}. The higher-twist LCDAs  are
not suppressed in many processes such as radiative leptonic $B$
meson decay $B \to \gamma\ell\nu $
\cite{Wang:2018wfj,Shen:2018abs}. In this paper we will
investigate the contribution from the full set the LCDAs of photon
up to twist-4 to the pion transition form factor using LCSRs
approach.

The outline of this paper is  as follow:  in Section
\ref{sect:higher-twist-photon} we present the analytic calculation
of the pion transition form factor from the higher twist photon
LCDAs within LCSRs framework. The numerical results and
discussions are given in section \ref{numerical}. The last section
is closing remark.

%********************************************************************************
\section{Power corrections from the hadronic structure of photon}
\label{sect:higher-twist-photon}
%********************************************************************************

All two-particle and three-particle LCDAs of  photon have been
defined and classified  up to twist-4, and the expressions of the
LCDAs have also been obtained through the conformal expansion in
the presence of the background field \cite{Ball:2002ps}.  To
evaluate the power suppressed contribution to the pion-photon form
factor due to the hadronic photon effect, the following
correlation function is employed
\begin{eqnarray}
G_{\mu}(p^{\prime}, q) &=& \int d^4 z \, e^{-i \, q \cdot z}  \,
\langle 0 | {\rm T} \left \{ j_{\mu, \perp}^{\rm em}(z),
j_{\pi}(0) \right \} | \gamma(p^{\prime}) \rangle  \,
\nonumber \\
&=& - g_{\rm em}^2 \, \epsilon_{\mu \nu \alpha \beta}^{\perp} \,
q^{\alpha} \, p^{\prime \beta} \, \epsilon^{\nu}(p^{\prime}) \,
G(p^2, Q^2) \,, \label{vacuum to photon correlator}
\end{eqnarray}
where pion interpolating current $j_{\pi}$ is defined by
\begin{eqnarray}
j_{\pi} = {1 \over \sqrt{2}} \, \left ( \bar u \, \gamma_5 \, u -
\bar d \, \gamma_5 \, d  \right )   \,.
\end{eqnarray}
The power counting rule for the external momenta
\begin{eqnarray}
| n \cdot p | \sim \bar n \cdot p \sim n \cdot p^{\prime} \sim
{\cal O}(\sqrt{Q^2}) \,,
\end{eqnarray}
will  be adopted to determine the perturbative matching
coefficient entering the factorization formula of
$G_{\mu}(p^{\prime}, q) $. Applying the standard definition for
the pion decay constant
\begin{eqnarray}
\langle 0 | j_{\pi}| \pi(p) \rangle = -i \, f_{\pi}  \,
\mu_{\pi}(\mu) \,, \qquad \mu_{\pi}(\mu) \equiv {m_{\pi}^2 \over
m_u(\mu) + m_d(\mu) } \,,
\end{eqnarray}
we can write down the hadronic dispersion relation of $G(p^2,
Q^2)$
\begin{eqnarray}
G(p^2, Q^2)  =  {f_{\pi}  \, \mu_{\pi}(\mu) \over m_{\pi}^2  - p^2
- i 0} \, F^{\rm NLP}_{\gamma^{\ast} \gamma \to \pi^0} (Q^2) +
\int_{s_0}^{\infty} \,ds \, {\rho^{h}(s, Q^2) \over s-p^2-i 0} \,.
\end{eqnarray}
The form factor $F^{\rm NLP}_{\gamma^{\ast} \gamma \to \pi^0}
(Q^2)$ will be extracted after the correlation function being
calculated by OPE in deep Euclidean region. Employing dispersion
relation, subtracting the continuum state contribution with the
help of quark hadron duality assumption, and performing Borel
transformation, the LCSRs for the subleading power contribution to
the $\pi^0 \gamma^{\ast} \gamma$ form factor are derived as
\begin{eqnarray}
F^{\rm 2PLT}_{\gamma^{\ast} \gamma \to \pi^0} (Q^2) &=& - {
\sqrt{2} \, \left ( Q_u^2 - Q_d^2 \right ) \over f_{\pi}  \,\,
\mu_{\pi}(\mu) \,\, Q^2} \, \chi(\mu) \, \langle \bar q q \rangle
(\mu) \,
\int_0^{s_0}  \, ds \, {\rm exp} \left[ - {s - m_{\pi}^2  \over M^2} \right ] \nonumber \\
&& \times \left [ \rho^{(0)}(s, Q^2) + {\alpha_s \, C_F \over 4\,
\pi} \,  \rho^{(1)}(s, Q^2)  \right ] + {\cal O}(\alpha_s^2)  \,.
\label{NLL resummation for the NLP effect}
\end{eqnarray}
where  the magnetic susceptibility of the quark condensate
$\chi(\mu)$ contains the dynamical information of the QCD vacuum,
and the spectral functions $\rho^{(0,1)}(s, Q^2)$ can be found in
\cite{Wang:2017ijn}.

Now we will proceed to investigate the contribution from higher
twist LCDAs of photon. Up to twist-4, the two-particle LCDAs of
photon are defined as
\begin{eqnarray}
&& \langle 0 |\bar q(x) [x, 0] \sigma_{\alpha \beta} \,\, q(0)|
\gamma(p) \rangle = i \, g_{\rm em} \, Q_q \,  \langle \bar q q
\rangle(\mu) \, (p_{\beta} \, \epsilon_{\alpha} - p_{\alpha} \,
\epsilon_{\beta}) \, \int_0^1 \, d z \, e^{i \, z \, p \cdot x} \,
\bigg [  \chi(\mu) \, \phi_{\gamma}(z, \mu)
 \nonumber \\
&& \,\,\,\,\,\,\,\,\,\,+ {x^2 \over 16} \, \mathbb{A}(z, \mu)
\bigg ] +{i \over 2} \, g_{\rm em} \, Q_q \, {\langle \bar q q
\rangle(\mu) \over p \cdot x} \, (x_{\beta} \, \epsilon_{\alpha} -
x_{\alpha} \, \epsilon_{\beta}) \,\nonumber
\int_0^1 \, d z \, e^{i \, z \, p \cdot x} \, h_{\gamma}(z, \mu) \,. \\
\nonumber && \langle 0 |\bar q(x) [x, 0] \gamma_{\alpha} \,\,
q(0)| \gamma(p) \rangle =  g_{\rm em} \, Q_q \, f_{3 \gamma}(\mu)
\, \epsilon_{\alpha} \, \int_0^1 \, d z \, e^{i \, z \, p \cdot x}
\, \psi_\gamma^{(v)}(z, \mu)  \,\\  && \langle 0 |\bar q(x) [x,
0]\gamma_{\alpha} \, \gamma_5 \,\, q(0)| \gamma(p) \rangle
={g_{\rm em}Q_q \, f_{3 \gamma}(\mu) \, \over 4} \,
\varepsilon_{\alpha \beta \rho \tau} \, p^{\rho} \, x^{\tau}
\,\epsilon^{\, \beta} \, \int_0^1 \, d z \, e^{i \, z \, p \cdot
x} \, \, \psi_\gamma^{(a)}(z, \mu)\,,\label{two
particle}\end{eqnarray} where $\psi_\gamma^{(v)}(z,
\mu),\psi_\gamma^{(a)}(z, \mu)$ are twist-3 and $\mathbb{A}(z,
\mu), h_{\gamma}(z, \mu)$ are twist-4.  Employing the light-cone
expansion of the $u,d$-quark propagator and keeping the
subleading-power contributions to the correlation function
(\ref{vacuum to photon correlator}) leads to
\begin{eqnarray}
G_{\mu}(p', q) & \supset & {1\over \sqrt{2}}\int {d^4 k \over (2
\pi)^4} \, \int d^4 x \, e^{i \, (k-q) \cdot x} \, {k^{\nu} \over
k^2 } \, \sum_{q=u,d}{\delta_q}Q_qg_{em}\langle0| \bar q(x) \,
\sigma_{\mu \nu} \, \gamma_5 \, q(0)| \gamma(p') \rangle
\,-\left(q\leftrightarrow -p\right)
 \nonumber \\
&=&{i\over 2\sqrt 2}\epsilon_{\mu\nu\rho\sigma}\int {d^4 k \over
(2 \pi)^4} {k^{\nu} \over k^2 }\, \int d^4 x \, e^{i \, (k-q)
\cdot x} \, \,\sum_{q=u,d}{\delta_q}Q_qg_{em} \langle 0| \bar q(x)
\, \sigma^{\rho\sigma} \, q(0)| \gamma(p') \rangle \,\nonumber
\\&-&\left(q\leftrightarrow -p\right),
\end{eqnarray}
where $\delta_u=1,\delta_d=-1$. The above equation indicates
that only twist-2 and twist-4  two-particle LCDAs can contribute
to pion transition form factor in the LCSRs approach, which is
different from the method based on TMD
factorization\cite{Shen2019}. Making use of the definitions in
Eq.(\ref{two particle}), it is straightforward to write down
\begin{eqnarray}
G_{\mu}^{\rm 2PHT}(p, q) &=& -{g_{\rm em}^2 \over 4 }\,
\epsilon_{\mu \nu \alpha \beta}^{\perp} \varepsilon^{\nu}\,
q^{\alpha} \, p^{\prime \beta} \,  {Q_u^2-Q_d^2\over \sqrt 2
Q^4}\langle\bar qq\rangle(\mu)\int_0^1du\left[{\mathbb{A}(u,
\mu)\over (\bar u+ur)^2}+{\mathbb{A}(u, \mu)\over ( u+r\bar
u)^2}\right],
\end{eqnarray}
where the contribution from $h_\gamma(z,\mu)$ vanishes due to the
anti-symmetric structure.  The resulting LCSRs for the
two-particle higher-twist hadronic photon corrections to the pion
transition form factors can be further derived as follows
\begin{eqnarray}
F^{\rm 2PHT}_{\gamma^{\ast} \gamma \to \pi^0} (Q^2) &=& - {
\sqrt{2} \, \left ( Q_u^2 - Q_d^2 \right ) \over 4f_{\pi}  \,\,
\mu_{\pi}(\mu) \,\,} \,  \langle \bar q q \rangle (\mu) \,\bigg\{
{1\over Q^2} \mathbb{A}(u_0)e^{-{s_0-m_\pi^2\over M^2}}\nonumber
\\ &+&\int_{u_0}^{1} \, {du\over u^2 }\, {1\over M^2}{\rm exp}
 \left[ - {\bar uQ^2 - um_{\pi}^2  \over uM^2} \right ] \mathbb{A}(u,
 \mu),\bigg\}
\label{2pht}
\end{eqnarray}
where $u_0=Q^2/(s_0+Q^2)$.

%%%%%%%%%%%
\begin{figure}
\begin{center}
\includegraphics[width=0.45\columnwidth]{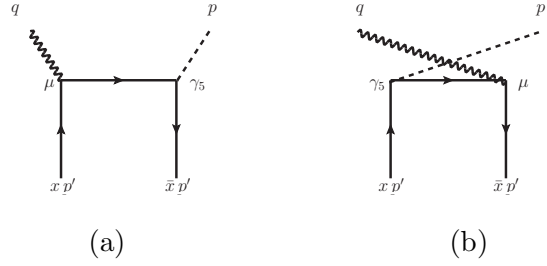}
\\
(a) \hspace{4 cm} (b) \vspace*{0.1cm} \caption{Diagrammatical
representation of the tree-level contribution to the QCD amplitude
$\widetilde{G}_{\mu}$ with the contribution from two-particle
photon LCDAs. } \label{fig:tree-level-pion-FF-LP}
\end{center}
\end{figure}
%%%%%%%%%%%
%%%%%%%%%%%
\begin{figure}
\begin{center}
\includegraphics[width=0.40 \columnwidth]{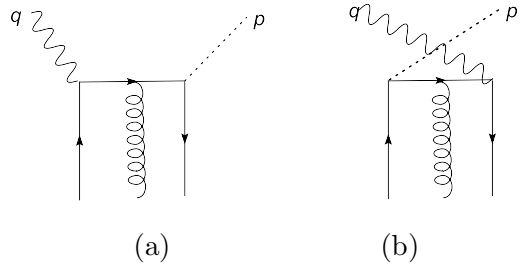} \\
\hspace{0.5cm}(a) \hspace{2cm} \hspace{0.5cm}(b) \vspace*{0.1cm}
\caption{Diagrammatical representation of the tree-level
contribution to the three-particle photon LCDAs. }
\label{fig:tree-level-pion-FF-NLP}
\end{center}
\end{figure}

To compute  higher-twist three-particle hadronic photon
corrections to the pion transition form factors, the definition of
three-particle photon LCDA is required. In the appendix we
collect the definition of three-particle twist-4 photon LCDAs for
an incoming photon state. Keeping the one-gluon/photon part for
the light-cone expansion of the quark propagator in the background
gluon/photon field
\begin{eqnarray}
 \langle 0|T\{{q}(x),\bar{q}(0)\}|0\rangle_{G}&\supset& i\int\frac{d^{4}k}{(2\pi)^{4}}
 e^{-ik\cdot x}\int_{0}^{1}du[\frac{ux_{\mu}\gamma_{\nu}}{k^{2}}
 -\frac{\not \! k\sigma_{\mu\nu}}{2k^4}]G^{\mu\nu}(ux)\nonumber
 \\ &+&
 ig_{em}Q_q\int^{\infty}_{0}\frac{d^{4}k}{(2\pi)^{4}}
 e^{-ik\cdot x}\int_{0}^{1}du[\frac{ux_{\mu}\gamma_{\nu}}{k^{2}}
 -\frac{\not \! k\sigma_{\mu\nu}}{2k^4}]F^{\mu\nu}(ux)
\end{eqnarray}where
 $ G^{\mu\nu}=i[D_{\mu},D_{\nu}]$. By evaluating Fig. \ref{fig:tree-level-pion-FF-NLP}, we obtain
\begin{eqnarray}
\Pi_{\mu}(p, q) & \supset &{1\over
2\sqrt{2}}g_{em}^2\sum_q\delta_qQ_q^2\epsilon_{\mu\alpha\rho\lambda}q^\alpha
\varepsilon^\rho p'^\lambda\langle \bar q q \rangle(\mu)\int_0^1 du\int
[{\cal D} \alpha_i]{1\over[ q-(\alpha_q +  \, \bar u \,
\alpha_g-1) \, p']^4}\nonumber \\
&\times& \rho^{\rm 3PHT}(\alpha_i, u, \mu)-\left(q\leftrightarrow
-p\right)
\end{eqnarray}
where
\begin{eqnarray}
\rho^{\rm 3PHT}(\alpha_i, u, \mu)&=&
2\{(2u-1)[T_1(\alpha_i)-T_2(\alpha_i)+T_3(\alpha_i)+T_4(\alpha_i)-\widetilde{S}(\alpha_i)
+T_{4\gamma}(\alpha_i)]\nonumber \\
&+&S(\alpha_i, \mu)+S_\gamma(\alpha_i, \mu)+T_2(\alpha_i,
\mu)-T_1(\alpha_i, \mu)\}
\end{eqnarray}
and the integration measure is defined as
\begin{eqnarray}
\int [{\cal D} \alpha_i] \equiv \int_0^1 d \alpha_q \, \int_0^1 d
\alpha_{\bar q} \, \int_0^1 d \alpha_g \, \delta \left (1-\alpha_q
- \alpha_{\bar q} -\alpha_g \right )\,.
\end{eqnarray}
Taking advantage of quark-hadron duality, we arrive at the LCSRs
of the contribution from three-particle photon LCDAs
\begin{eqnarray}
F^{\rm 3PHT}_{\gamma^{\ast} \gamma \to \pi^0} (Q^2) &=& - {
\sqrt{2} \, \left ( Q_u^2 - Q_d^2 \right ) \over 2f_{\pi}  \,\,
\mu_{\pi}(\mu) \,\,} \,  \langle \bar q q \rangle (\mu){1\over
Q^2} \,
\bigg\{\int_0^{s_0/(s_0+Q^2)}d\alpha_q\int_{s_0/(s_0+Q^2)-\alpha_q}^{1-\alpha_q}{d\alpha_g\over
\alpha_g}
\nonumber \\
&\times& \rho^{\rm 3PHT}(\alpha_q,\alpha_g,\alpha_{\bar
q}=1-\alpha_q-\alpha_g,u_{s_0},\mu)e^{-{s_0-m_\pi^2\over M^2}}\nonumber \\
&+&{1\over M^2}\int_0^{s_0}dse^{-{s-m_\pi^2\over
M^2}}\int_0^{s/(s+Q^2)}d\alpha_q\int_{s/(s+Q^2)-\alpha_q}^{1-\alpha_q}{d\alpha_g\over
\alpha_g}
\nonumber \\
&\times& \rho^{3PHT}(\alpha_q,\alpha_g,\alpha_{\bar
q}=1-\alpha_q-\alpha_g,u_s,\mu)\bigg\} \label{3pht}
\end{eqnarray}
where $u_s=[s/(s+Q^2)-\alpha_q]/\alpha_g$. The overall higher-twist photon LCDAs contribution is written by
\begin{eqnarray}
F^{\rm HT}_{\gamma^{\ast} \gamma \to \pi^0} (Q^2)=F^{\rm
2PHT}_{\gamma^{\ast} \gamma \to \pi^0} (Q^2)+F^{\rm
3PHT}_{\gamma^{\ast} \gamma \to \pi^0} (Q^2).
\end{eqnarray}
Now we discuss the power behavior of our results. The power
counting scheme for the  sum rule parameters are given below:
\begin{eqnarray}
s_0 \sim M^2 \sim {\cal O} (\Lambda^2) \,, \qquad \bar u_0 \sim
{\cal O} (\Lambda^2/Q^2).  \label{scaling}
\end{eqnarray}
Employing Eq.(\ref{scaling}), one can obtain that the contribution
from leading twist LCDA of photon is suppressed by a factor
$\Lambda^2/Q^2$ \cite{Wang:2017ijn} compared with LP contribution.
The higher twist contributions are conjectured to be also
suppressed by {only} one power of $\Lambda^2/Q^2$ due to the
absent correspondence between the twist counting and the
large-momentum expansion \cite{Agaev:2010aq}.  For the
contribution from two-particle twist-4 LCDAs of photon, the result
in Eq.(\ref{2pht}) is suppressed by $\Lambda^4/Q^4$ compared with
LP contribution as the power of twist-4 photon LCDAs is suppressed
with respect to leading twist one. While for the contribution from
three-particle twist-4 LCDAs in Eq.(\ref{3pht}), the scaling of
$\alpha_q$ is ${\cal O}(\Lambda^2/Q^2)$, and $\alpha_g$ is ${\cal
O}(1)$. Although there is an overall factor $1/Q^2$, the result is
only suppressed by $\Lambda^2/Q^2$ for the spectral function
$\rho^{\rm 3PHT}$ is not suppressed at endpoint region. This
result confirms the conjecture in \cite{Agaev:2010aq}.

%********************************************************************************
\section{Numerical analysis}
\label{numerical}
%********************************************************************************

In the following we explore the phenomenological consequences of
the hadronic photon correction to the pion-photon form factor, and
the most important input is the LCDAs of photon. The models of
twist-4 LCDAs of photon used in this paper are written by
\begin{eqnarray}
\mathbb{A}(z, \mu) &=& 40 \, z^2 \, \bar z^2  \left [3 \,
\kappa(\mu) - \kappa^{+}(\mu) + 1 \right ] + 8 \, \left [
\zeta_2^{+} (\mu) - 3 \, \zeta_2(\mu) \right ] \,
\big [ z \, \bar z \, (2 + 13 \,z \, \bar z )  \nonumber \\
&& + \, 2\, z^3 \, (10 - 15\, z + 6 \, z^2) \, \ln z
+ 2 \, \bar z^3 \, (10 - 15 \,\bar  z + 6 \, \bar z^2)  \, \ln \bar z \big  ] \,, \nonumber \\
h_{\gamma}(z, \mu) &=& -10 \, \left (1 + 2 \, \kappa^{+}(\mu)
\right) \,
C_{2}^{1/2}(2 \, z -1)  \,, \nonumber \\
S(\alpha_i, \mu) &=& 30 \, \alpha_g^2 \, \bigg \{ \left
(\kappa(\mu) + \kappa^{+}(\mu) \right ) \, (1-\alpha_g)
+ (\zeta_1 + \zeta_1^{+}) (1 -\alpha_g) (1 -2 \, \alpha_g) \nonumber \\
&& + \, \zeta_2(\mu) \,\left [ 3 \, (\alpha_{\bar q} - \alpha_q)^2
- \alpha_g \, (1- \alpha_g) \right ] \bigg \}  \,, \nonumber \\
\widetilde{S}(\alpha_i, \mu) &=& - 30 \, \alpha_g^2 \, \bigg \{
\left (\kappa(\mu) - \kappa^{+}(\mu) \right ) \, (1-\alpha_g)
+ (\zeta_1 - \zeta_1^{+}) (1 -\alpha_g) (1 -2 \, \alpha_g) \nonumber \\
&& + \, \zeta_2(\mu) \,\left [ 3 \, (\alpha_{\bar q} - \alpha_q)^2
- \alpha_g \, (1- \alpha_g) \right ] \bigg \} \,, \nonumber \\
S_{\gamma}(\alpha_i, \mu) &=& 60 \, \alpha_g^2 \, (\alpha_q +
\alpha_{\bar q} ) \,
\left [ 4 - 7 \, (\alpha_{\bar q} + \alpha_q )  \right ] \,, \nonumber \\
T_{1}(\alpha_i, \mu) &=& - 120 \, \left (3 \, \zeta_2(\mu) +
\zeta_2^{+}(\mu) \right ) \,
\left ( \alpha_{\bar q} - \alpha_q \right )  \, \alpha_{\bar q} \, \alpha_q \, \alpha_g \,, \nonumber \\
T_{2}(\alpha_i, \mu) &=& 30 \, \alpha_g^2 \, (\alpha_{\bar q} -
\alpha_q )
  \left [  \left (\kappa(\mu) - \kappa^{+}(\mu) \right )
+ \left (\zeta_1(\mu) - \zeta_1^{+}(\mu) \right )(1-2 \, \alpha_g)
+ \zeta_2(\mu) \, (3 -4 \, \alpha_g)\right ] \,, \nonumber \\
T_{3}(\alpha_i, \mu) &=& -120 \, \left (3 \, \zeta_2(\mu) -
\zeta_2^{+}(\mu) \right ) \,
(\alpha_{\bar q} - \alpha_q ) \,\alpha_{\bar q} \, \alpha_q \, \alpha_g \,, \nonumber \\
T_{4}(\alpha_i, \mu) &=& 30 \, \alpha_g^2  \, (\alpha_{\bar q}
-\alpha_q)\, \left [  \left (\kappa(\mu) + \kappa^{+}(\mu) \right
) + \left (\zeta_1(\mu) + \zeta_1^{+}(\mu) \right )(1-2 \,
\alpha_g)
+ \zeta_2(\mu) \, (3 - 4 \, \alpha_g)\right ]  \,, \nonumber \\
T_{4}^{\gamma}(\alpha_i, \mu) &=& 60 \, \alpha_g^2 \, (\alpha_q -
\alpha_{\bar q} ) \, \left [ 4 - 7 \, (\alpha_{\bar q} + \alpha_q
)  \right ]  \,.
\end{eqnarray}
In the above equations,   the conformal expansion of the photon
LCDAs have been truncated up to the next-to-leading conformal
spin. Due to the Ferrara-Grillo-Parisi-Gatto theorem
\cite{Ferrara:1972xq}, these  parameters  satisfy the following
relations
\begin{eqnarray}
\zeta_1(\mu) + 11 \,\zeta_2(\mu) - 2 \, \zeta_2^{+}(\mu) = {7
\over 2} \,.
\end{eqnarray}
The scale evolution of the nonperturbative parameters  is given by
\begin{eqnarray}
\kappa^{+}(\mu)= \left (  {\alpha_s (\mu) \over \alpha_s (\mu_0)}
\right ) ^{\left ( \gamma^{+} - \gamma_{q \bar q} \right )  /
\beta_0} \, \kappa^{+}(\mu_0)\,, & \qquad & \kappa(\mu)= \left (
{\alpha_s (\mu) \over \alpha_s (\mu_0)} \right ) ^{\left (
\gamma^{-} - \gamma_{q \bar q} \right )  / \beta_0} \,
\kappa(\mu_0) \,, \nonumber \\
\zeta_1(\mu)= \left (  {\alpha_s (\mu) \over \alpha_s (\mu_0)}
\right ) ^{\left ( \gamma_{Q^{(1)}} - \gamma_{q \bar q} \right )
/ \beta_0} \, \zeta_1(\mu_0)\,,  & \qquad & \zeta_1^{+}(\mu)=
\left (  {\alpha_s (\mu) \over \alpha_s (\mu_0)} \right ) ^{\left
( \gamma_{Q^{(5)}} - \gamma_{q \bar q} \right )  / \beta_0} \,
\zeta_1^{+}(\mu_0)\,, \nonumber \\
\zeta_2^{+}(\mu)= \left (  {\alpha_s (\mu) \over \alpha_s (\mu_0)}
\right ) ^{\left ( \gamma_{Q^{(3)}} - \gamma_{q \bar q} \right )
/ \beta_0} \, \zeta_2^{+}(\mu_0)\,, \label{evolution of twist-four
parameters}
\end{eqnarray}
where the anomalous dimensions  at one loop read
\cite{Ball:2002ps}
\begin{eqnarray}
\gamma^{+}= 3 \, C_A - {5 \over 3} \, C_F \,, & \qquad &
\gamma^{-}= 4 \, C_A - 3 \, C_F \,,\,\,\,\,\,\,\,\,\gamma_{q \bar q} =-3 \, C_F \,, \nonumber \\
  \gamma_{Q^{(1)}} = {11
\over 2} \, C_A - 3 \, C_F \,,& \qquad & \gamma_{Q^{(3)}} = {13
\over 3} \, C_F \,, \,\,\,\,\,\,\,\,\gamma_{Q^{(5)}} = 5 \, C_A -
{8 \over 3} \, C_F \,.
\end{eqnarray}
Numerical values of the input parameters entering the photon LCDAs
up to twist-4 are collected in Table \ref{tab of parameters for
photon DAs}, where for the estimates of the twist-4 parameters
from QCD sum rules \cite{Balitsky:1989ry} 100 \% uncertainties are
assigned.

%%%%%%%%%%%%%%%%%%%%%%%%%%%%%%%%%%%%
\begin{table}[t]
\begin{center}
\begin{tabular}{|c|c|c|c|c|c|c|c|}
  \hline
  \hline
&&&&&&& \\[-3.5mm]
$\chi(\mu_0) \, $ & $\langle \bar q q \rangle (\mu_0) $
 & $b_2(\mu_0)$ & $\kappa(\mu_0)$ &
$\kappa^{+}(\mu_0)$ & $\zeta_{1}(\mu_0)$ & $\zeta_{1}^{+}(\mu_0)$
&
$\zeta_{2}^{+}(\mu_0)$  \\
    &&&&&&& \\[-1mm]
  \hline
    &&&&&&& \\[-1mm]
  $(3.15 \pm 0.3) \, {\rm GeV}^{-2}$ &$-(246^{+28}_{-19} \, {\rm MeV})^3$ & $0.07
\pm 0.07$ & $0.2 \pm 0.2$ &$0$&
$0.4 \pm 0.4$ & $0$  & $0$ \\
&&&&&&& \\
\hline \hline
\end{tabular}
\end{center}
\caption{Numerical values of the nonperturbative parameters
entering the photon LCDAs at the  scale $\mu_0= 1.0 \, {\rm GeV}$
\cite{Ball:2002ps,Duplancic:2008ix}.} \label{tab of parameters for
photon DAs}
\end{table}
%%%%%%%%%%%%%%%%%%%%%%%%
%%%%%%%%%%%%%%%%%%%%%%%%%%%%%%%%%%%%
\begin{table}[t]
\begin{center}
\begin{tabular}{|c|c|c|c|c|c|}
  \hline
  \hline
Models &CZ  & BMS &   KMOW & Holographic& Platykurtic  \\
    [-1mm]
  \hline
      $a_2$(1GeV)  &0.5&$0.20^{+0.07}_{-0.08}$&$0.17\pm0.08$& 0.15&0.08 \\[-1mm]
  \hline
   $a_4$(1GeV) &0&$-0.15^{+0.10}_{-0.09}$&$0.06\pm0.10$&0.06&-0.02 \\
\hline \hline
\end{tabular}
\end{center}
\caption{The numerical values of Gegenbauer momemts $a_2$ and
$a_4$ in leading twist pion LCDA.} \label{tab Gegenbauer moments}
\end{table}
%%%%%%%%%%%%%%%%%%%%%%%%

\begin{figure}
\begin{center}
\includegraphics[width=0.45\columnwidth]{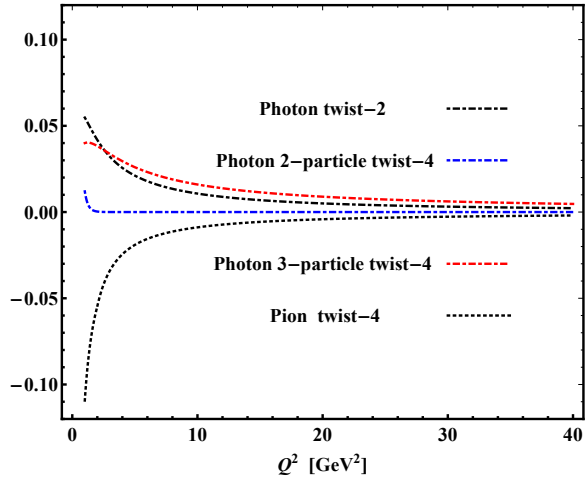} \\
\vspace*{0.1cm} \caption{Comparison of the power suppressed
contribution to pion-photon form factor $Q^2F_{\gamma^{\ast}
\gamma \to \pi^{0}}(Q^2)$ from different sources . }
\label{fig:comparison}
\end{center}
\end{figure}

Now we are in the position to investigate the phenomenological
significance of the contribution from higher twist photon LCDAs.
For the factorization scale in the evaluation of the contribution
of higher-twist photon LCDAs, we will take the value $\mu^2=
\langle x \rangle \, M^2 + \langle \bar x \rangle \, Q^2$ as
widely employed in the sum rule calculations \cite{Agaev:2010aq}.
The Borel mass $M^2$ and the threshold parameter $s_0$ can be
determined by applying the standard strategies described in
\cite{Wang:2015vgv,Wang:2017jow},
\begin{eqnarray}
M^2= (1.25 \pm 0.50) \, {\rm GeV^2}   \,, \qquad s_0= (0.70 \pm
0.20) \, {\rm GeV^2}   \,,\label{borel}
\end{eqnarray}
where the variation ranges of these parameters are set to be large
to allow sufficient theoretical uncertainty. It has been checked
that the Borel mass and  threshold parameter dependence of the
contribution of higher-twist photon LCDAs is mild in the intervals
in Eq.(\ref{borel}). In Fig. \ref{fig:comparison} the $Q^2$
dependence of the relevant power suppressed contributions is
presented. Compared with the contribution from leading-twist
photon LCDA, the two-particle twist-4 contribution is obviously
suppressed  as the curve declines more quickly and approaches zero
at large $Q^2$. While for the contribution from three-particle
twist-4 LCDAs of photon, the result is comparable with that from
leading twist photon LCDA, as they are at the same power.  As
mentioned in \cite{Wang:2017ijn}, there exists strong cancellation
effect between the contribution from leading twist photon LCDA and
the twist-4 pion LCDA , thus the overall power correction is
mainly from the contribution from twist-4 LCDAs of photon.

\begin{figure}
\begin{center}
\includegraphics[width=0.45 \columnwidth]{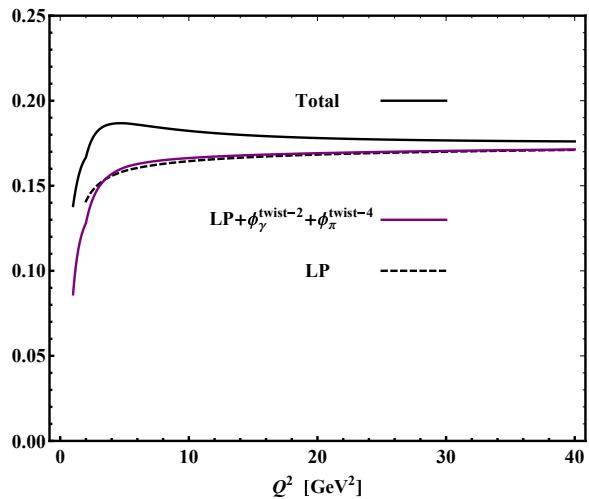} \\
\vspace*{0.1cm} \caption{Total result of the pion-photon form
factors $Q^2F_{\gamma^{\ast} \gamma \to \pi^{0}}(Q^2)$ after
including power corrections. } \label{fig:total}
\end{center}
\end{figure}

To obtain the total result of the photon-pion form factor, we will
need to specify the non-perturbative models for the twist-2 pion
LCDA. In general it is expanded in terms of Gegenbauer polynomials
\begin{eqnarray}
\phi_{\pi}(x ,\mu) = 6 \, x \, \bar x \, \sum_{n=0}^{\infty} \,
a_n(\mu) \, C_n^{3/2}(2 x-1)\,, \label{Gegenbauer expansion of
pion DA}
\end{eqnarray} where the Gegenbauer moments $a_n$ can be determined by the
calculation with QCD sum rules or lattice simulation, or by
fitting the experimental data. Following \cite{Wang:2017ijn}, we
take advantage of the the Chernyak- Zhitnitsky (CZ)
model\cite{Chernyak:1981zz}, the Bakulev-Mikhailov-Stefanis (BMS)
model\cite{Bakulev:2001pa}, the platykurtic model
(PK)\cite{Stefanis:2014nla}, the KMOW
model\cite{Khodjamirian:2011ub},  and the holographic
model\cite{Brodsky:2007hb} for comparison. The Gegenbauer
coefficients in the BMS model and the PK model are computed from
the QCD sum rules with non-local condensates, the first and second
nontrivial Gegenbauer moments of the KMOW model are determined by
comparing the LCSR predictions for the pion electromagnetic form
factor with the experimental data at intermediate-$Q^2$, and the
holographic model of the twist-2 pion LCDA is motivated by the
AdS/QCD correspondence. We collect the values of the Gegenbauer
moments in different models in Table. \ref{tab Gegenbauer
moments}. The total results including power suppressed
contributions are shown in Fig. \ref{fig:total}, where the BMS
model is employed. It can be seen that the higher power photon
LCDAs manifestly modify the LP result especially at ``small" $Q^2$
region. We note that the photon-LCSRs employed in this paper is
valid when $Q^2\gg2GeV^2$, thus the prediction of
$F_{\gamma^{\ast} \gamma \to \pi^0} (Q^2)$ should not be taken
serious below 2 GeV$^2$.
%%%%%%%%%%%
%%%%%%%%%%%
%%%%%%%%%%%
%%%%%%%%%%%
\begin{figure}
\begin{center}
\includegraphics[width=0.45 \columnwidth]{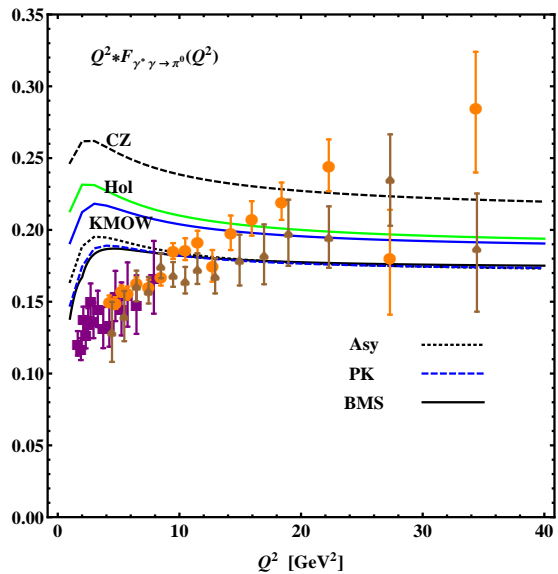} \\
\vspace*{0.1cm} \caption{Total result of the pion-photon form
factors $Q^2F_{\gamma^{\ast} \gamma \to \pi^{0}}(Q^2)$ with
different models of leading twist pion LCDA. Points from CLEO
\cite{Gronberg:1997fj} (purple squares), BaBar
\cite{Aubert:2009mc} (orange circles) and Belle
\cite{Uehara:2012ag} (brown spades) are  displayed here.}
\label{fig:models}
\end{center}
\end{figure}
%$$$$$$$$$$$$$$$$
The model dependence of pion-photon form factor on the leading
twist pion LCDA is displayed in Fig. \ref{fig:models}. As the
contribution from  higher twist photon LCDA enhances the form
factors significantly, the prediction from every models cannot
match the experimental data at $Q^2<10GeV^2$. This result is
inconsistent with the predictions from dispersion
approach\cite{Bakulev:2002uc,Stefanis:2012yw,Bakulev:2011rp,Bakulev:2012nh,Mikhailov:2016klg},
where the BMS and PK models of pion LCDA work well. This
discrepancy is not a surprise because the power suppressed
contributions considered in both approaches are not from a
systematic study based on the effective theory, and what is
omitted is not clear. Our result indicates that there exist
significant power suppressed contributions, and they should not be
neglected in the phenomenological studies. Meanwhile, we cannot
draw the conclusion that the models mentioned in this paper should
be ruled out, because in our study the QCD corrections are not
included, and contributions from the pion and photon LCDA with
twist higher than 4 are not considered, let alone the unknown
power suppressed contributions. Thus in the present paper we aim
at sheding light on the importance of the power corrections, and
more efforts must be devoted to the study on power suppressed
contributions to obtain more accurate prediction.

\begin{figure}
\begin{center}
\includegraphics[width=0.4
\columnwidth]{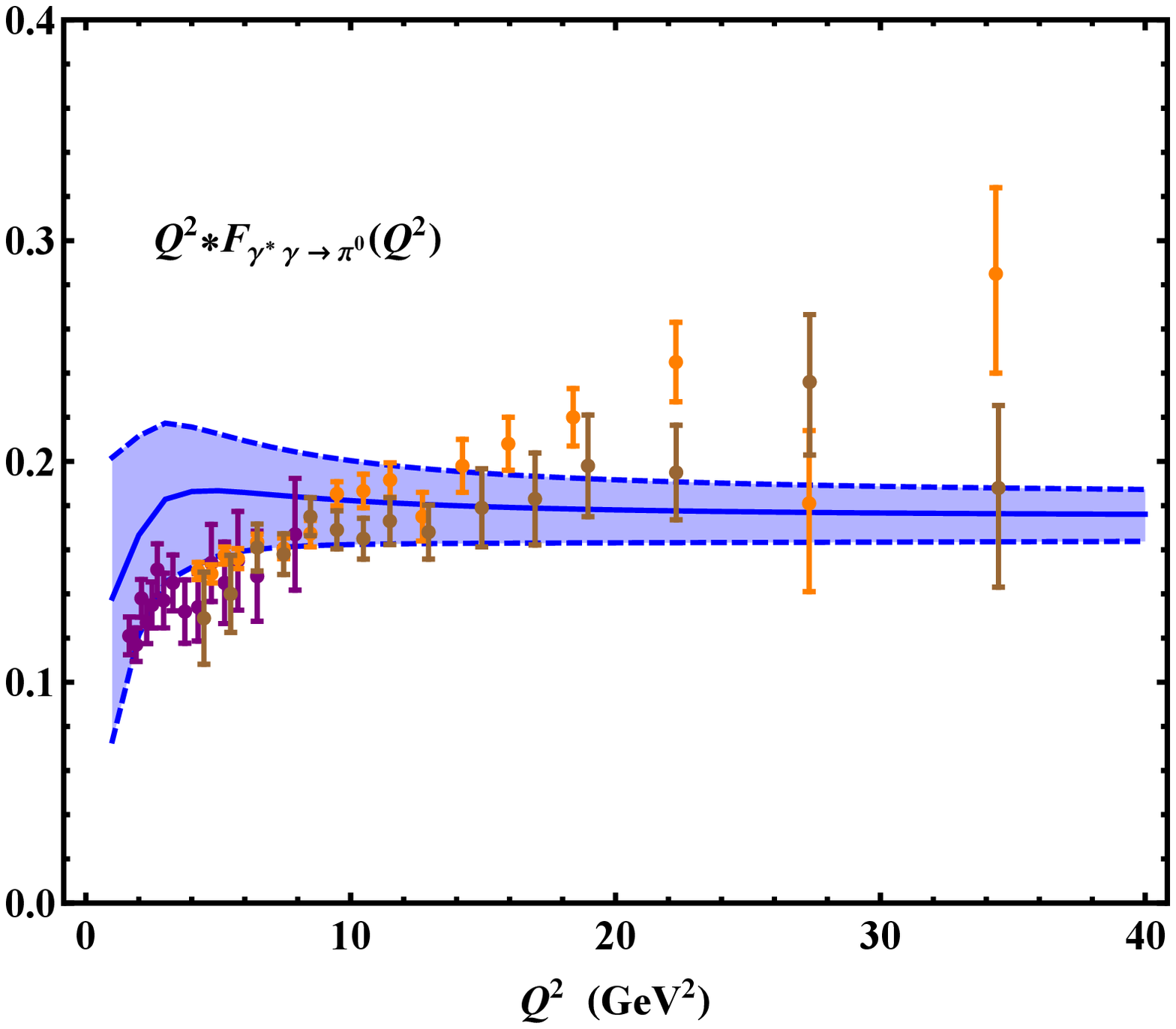}\hspace{-2.5cm}(a)\\\vspace{0.5cm}
\includegraphics[width=0.35 \columnwidth]{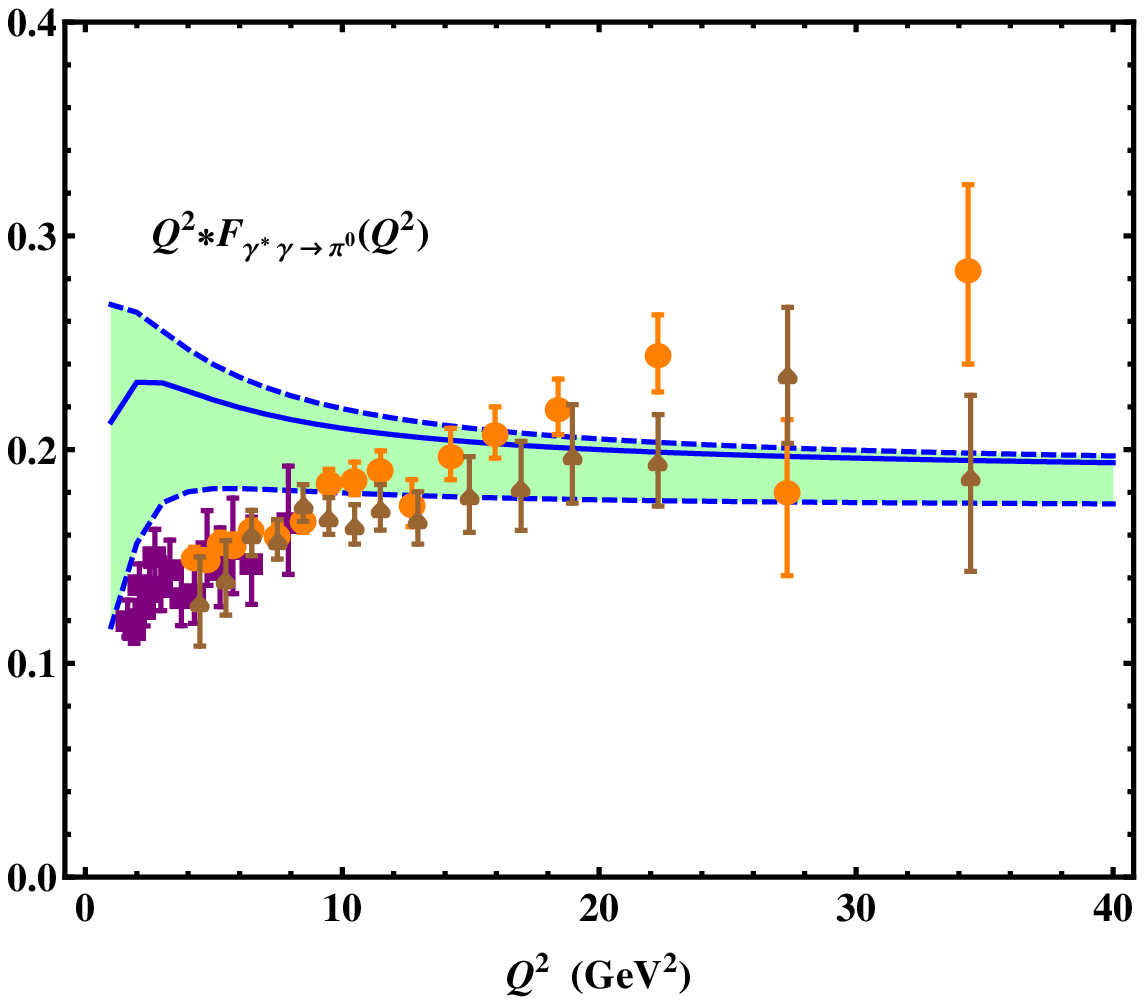}\hspace{-2.5cm}(b)\hspace{2cm}
\includegraphics[width=0.35 \columnwidth]{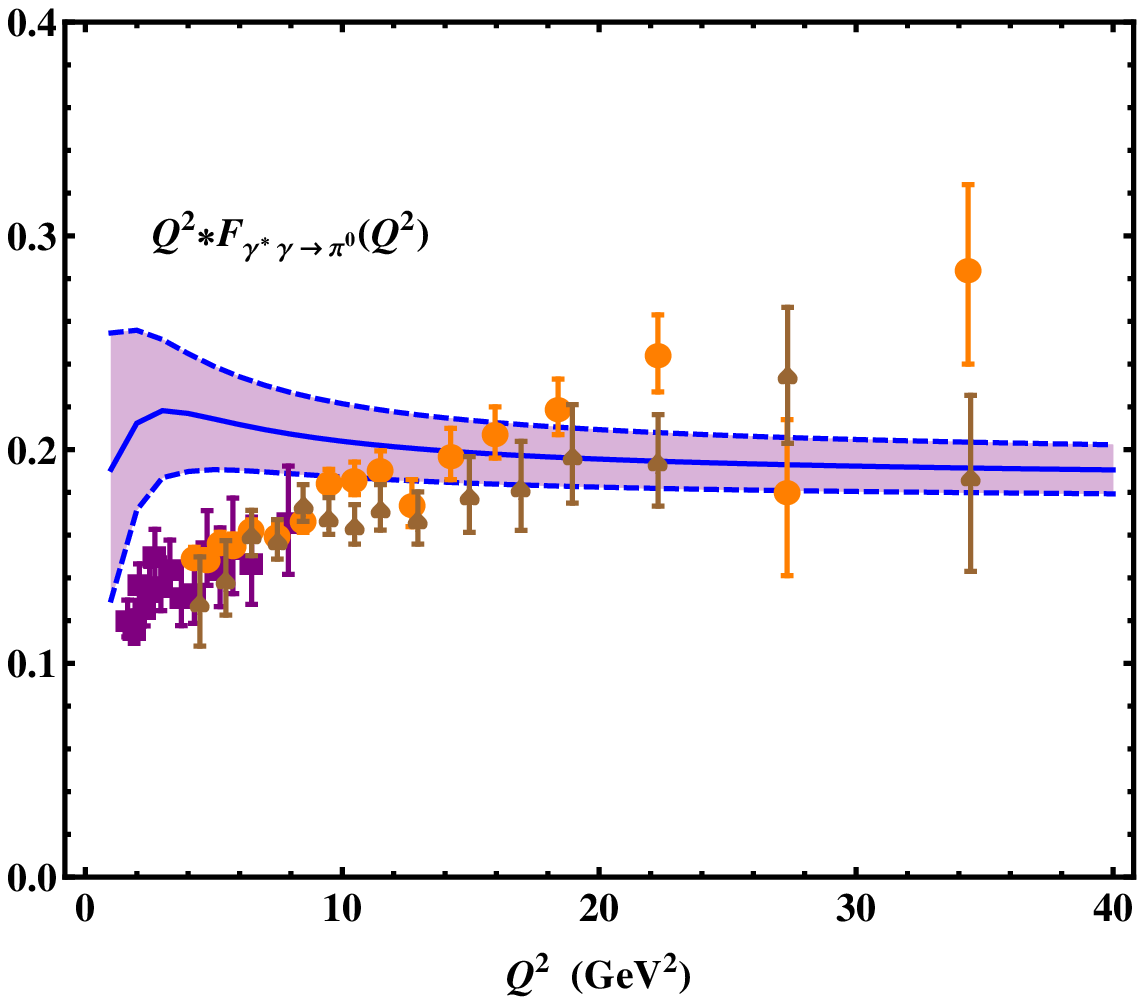}\hspace{-2.5cm}(c) \\
\vspace*{0.1cm} \caption{Comparison between the theoretical
predictions in this paper and the experimental data. Points from
CLEO \cite{Gronberg:1997fj} (purple squares), BaBar
\cite{Aubert:2009mc} (orange circles) and Belle
\cite{Uehara:2012ag} (brown spades) are  displayed here. Diagram
(a), (b) and (c) in Fig. \ref{fig:exp} are corresponding to the
BMS model, holographic model and KMOW model of pion LCDA are
employed respectively. } \label{fig:exp}
\end{center}
\end{figure}

We present our final predictions for $Q^2F_{\gamma^{\ast} \gamma
\to \pi^{0}}(Q^2)$ with both LP contribution and power corrections
included in Fig. \ref{fig:exp}, where the combined theory
uncertainties are due to the variations of the input parameters
$a_2,a_4$  of pion LCDA, $\xi, \langle \bar qq\rangle, b_2$ in
twist-2 photon LCDAs, $\kappa, \zeta_1,\zeta_2$ in twist-4 photon
LCDAs, quark mass, and factorization scale, etc. Diagram (a), (b)
and (c) in Fig. \ref{fig:exp} are corresponding to the BMS model,
holographic model and KMOW model of pion LCDA respectively.
 Among all the parameters, the most important
uncertainty comes from the shape parameters $a_2,a_4$ of leading
twist pion LCDA, which means the pion transition form factor is
still sensitive to the Gegenbauer moments of leading twist pion
LCDA after the power suppressed contributions considered. Thus the
photon-pion transition process provides a good platform to
determine the parameters in the LCDAs of pion, which can also be
compared with the future lattice simulation with the help of quasi
parton distribution amplitude \cite{Liu:2018tox,Wang:2017qyg}.

%%%%%%%%%%%

\section{Closing remark}

In this paper we performed a study on the power suppressed
contributions from higher-twist LCDAs of photon within the LCSRs.
The twist-3 LCDAs cannot contribute for their Lorentz structures,
thus the contributions from two-particle and three-particle
twist-4 LCDAs of photon are considered in this work. According to
the power analysis, the three-particle twist-4 contribution is not
suppressed compared with the leading twist photon LCDA result, so
that the power corrections considered in this work can give rise
to sizable contribution, especially at ``low" $Q^2$ region. In
addition, there exists strong cancellation between the
contribution from leading twist photon LCDA and the twist-4 pion
LCDA, and the importance of the twist-4 photon LCDAs is further
highlighted. The numerical result also confirms that after
including power corrections, the predicted $Q^2F_{\gamma^{\ast}
\gamma \to \pi^{0}}(Q^2)$ is significantly enhanced especially at
at ``low" $Q^2$ region, thus the power suppressed contributions
should be included in the determination of the Gegenbauer moments
of pion LCDAs. Note that for the higher-twist photon LCDAs
contribution, we only presented a tree level calculation, the NLO
QCD corrections which might modify the current result to some
extent and stablize the factorization scale dependence are not
considered. Furthermore, the other power suppressed contributions
are also absent in the present study,  a more systematic study
based on effective theory is necessary for a thorough
understanding of the NLP corrections to the pion transition form
factor, which can be checked by the (potentially) more accurate
experimental measurements at the BEPCII collider and the SuperKEKB
accelerator.

\subsection*{Acknowledgements}

We thank S. V.  Mikhailov and N. G. Stefanis for valuable
comments. This work is supported in part by the National Natural
Science Foundation of China (NSFC) with Grant No. 11521505 and
11621131001. CDL would like to express a special thanks to the
Mainz Institute for Theoretical Physics (MITP) for its hospitality
and Support.

%%%%%%%%%%%%%%%%%%%%%%%%%%%%%%%%%%%%%%%%%%%%%%%%%%%%%%%%

\appendix
\section{Definition of three-particle twist-4 LCDAs of photon}

%%%%%%%%%%%%%%%%%%%%%%%%%%%%%%%%%%%%%%%%%%%%%%%%%%%%%%%%

In the following, we present the definition of the three-particle
photon LCDAs up to twist-4.
\begin{eqnarray}
&& \langle 0 |\bar q(x) g_s \, G_{\alpha \beta}(u \, x) \, \, q(0)|\gamma(p) \rangle \nonumber \\
&& =  i \, g_{\rm em} \, Q_q \,  \langle \bar q q \rangle(\mu) \,
(p_{\beta} \, \epsilon_{\alpha} - p_{\alpha} \, \epsilon_{\beta})
\, \int [{\cal D} \alpha_i] \, e^{i \, (\alpha_q + \, \bar u
\, \alpha_g-1) \, p \cdot x}\, S(\alpha_i, \mu)
 \\
&& \langle 0 |\bar q(x)  g_s \, \widetilde{G}_{\alpha \beta}( u \,
x) \,
i \, \gamma_5 \,\,  q(0)|\gamma(p)\rangle  \nonumber \\
&& =  i \, g_{\rm em} \, Q_q \,  \langle \bar q q \rangle(\mu) \,
(p_{\beta} \, \epsilon_{\alpha} - p_{\alpha} \, \epsilon_{\beta})
\, \int [{ D} \alpha_i] \,  e^{i \, (\alpha_q + \, \bar u \,
\alpha_g-1) \, p \cdot x}\, \widetilde{S}(\alpha_i, \mu)
 \\
&& \langle  0|\bar q(x) g_s \, \widetilde{G}_{\alpha \beta}(u \,
x) \,
\gamma_{\rho} \, \gamma_5 \,\,  q(0)| \gamma(p) \rangle \nonumber \\
&& = - g_{\rm em} \, Q_q \, f_{3 \gamma}(\mu) \, p_{\rho} \,
(p_{\beta} \, \epsilon_{\alpha} - p_{\alpha} \, \epsilon_{\beta})
\, \int [{\cal D} \alpha_i] \, e^{i \, (\alpha_q + \, \bar u
\, \alpha_g-1) \, p \cdot x} \, A(\alpha_i, \mu)
 \\
&& \langle 0 |\bar q(x)  g_s \, G_{\alpha \beta}(u \, x) \,
i \, \gamma_{\rho} \,\,  q(0)|\gamma(p) \rangle \nonumber \\
&& =  g_{\rm em} \, Q_q \, f_{3 \gamma}(\mu) \,  p_{\rho} \,
(p_{\beta} \, \epsilon_{\alpha} - p_{\alpha} \, \epsilon_{\beta})
\, \int [{\cal D} \alpha_i] \, e^{i \, (\alpha_q + \, \bar u
\, \alpha_g-1) \, p \cdot x}\, V(\alpha_i, \mu)
 \\
&& \langle 0 |\bar q(x)  g_{\rm em} \, Q_q \,  F_{\alpha \beta}(u \, x) \, \, q(0)| \gamma(p) \rangle \nonumber \\
&& =  i \, g_{\rm em} \, Q_q \,  \langle \bar q q \rangle(\mu) \,
(p_{\beta} \, \epsilon_{\alpha} - p_{\alpha} \, \epsilon_{\beta})
\, \int [{\cal D} \alpha_i] \, e^{i \, (\alpha_q + \, \bar u
\, \alpha_g-1) \, p \cdot x} \, S_{\gamma}(\alpha_i, \mu) \,.  \\
\nonumber \\
&& \langle 0 |\bar q(x)  \,\, \sigma_{\rho \tau} \,\, g_s \,
G_{\alpha \beta}(u \, x)
\,\,  q(0)| \gamma(p)\rangle \nonumber \\
&& = - \, g_{\rm em} \, Q_q \,\langle \bar q q \rangle(\mu) \,
\left [p_{\rho} \, \epsilon_{\alpha} \, g_{\tau \beta}^{\perp} -
p_{\tau} \, \epsilon_{\alpha} \, g_{\rho \beta}^{\perp} - (\alpha
\leftrightarrow \beta) \right ]  \, \int [{\cal D} \alpha_i]
\,e^{i \, (\alpha_q + \, \bar u
\, \alpha_g-1) \, p \cdot x} \, T_{1}(\alpha_i, \mu)  \nonumber \\
&& \hspace{0.4 cm} - \, g_{\rm em} \, Q_q \,\langle \bar q q
\rangle(\mu) \, \left [p_{\alpha} \, \epsilon_{\rho} \, g_{\tau
\beta}^{\perp} - p_{\beta} \, \epsilon_{\rho} \, g_{\tau
\alpha}^{\perp} - (\rho \leftrightarrow \tau) \right ]  \, \int
[{\cal D} \alpha_i] \, e^{i \, (\alpha_q + \, \bar u
\, \alpha_g-1) \, p \cdot x}\, T_{2}(\alpha_i, \mu) \nonumber \\
&& \hspace{0.4 cm} - \, g_{\rm em} \, Q_q \,\langle \bar q q
\rangle(\mu) \, \frac{(p_{\alpha} \, x_{\beta} - p_{\beta} \,
x_{\alpha} ) (p_{\rho} \, \epsilon_{\tau} - p_{\tau}
\,\epsilon_{\rho})} {p \cdot x} \, \int [{\cal D} \alpha_i] \,
e^{i \, (\alpha_q + \, \bar u \, \alpha_g-1) \, p \cdot x} \,
 T_{3}(\alpha_i, \mu) \nonumber \\
&& \hspace{0.4 cm} - \, g_{\rm em} \, Q_q \,\langle \bar q q
\rangle(\mu) \, \frac{(p_{\rho} \, x_{\tau} - p_{\tau} \, x_{\rho}
) (p_{\alpha} \, \epsilon_{\beta} - p_{\beta}
\,\epsilon_{\alpha})} {p \cdot x} \, \int [{\cal D} \alpha_i] \,
e^{i \, (\alpha_q + \, \bar u \, \alpha_g-1) \, p \cdot x} \,
 T_{4}(\alpha_i, \mu) \,. \\
 \nonumber \\
&& \langle 0 |\bar q(x) \sigma_{\rho \tau} \,\, g_{\rm em} \, Q_q
\, F_{\alpha \beta}(u \, x)
\,\,  q(0)| \gamma(p) \rangle \nonumber \\
&& = - \, g_{\rm em} \, Q_q \,\langle \bar q q \rangle(\mu) \,
\frac{(p_{\rho} \, x_{\tau} - p_{\tau} \, x_{\rho} ) (p_{\alpha}
\, \epsilon_{\beta} - p_{\beta} \,\epsilon_{\alpha})} {p \cdot x}
\, \int [{\cal D} \alpha_i] \, e^{i \, (\alpha_q + \, \bar u
\, \alpha_g-1) \, p \cdot x}\,
 T_{4}^{\gamma}(\alpha_i, \mu)  + ... \,
\end{eqnarray}
Note that we have employed the following notations for the dual
field strength tensor and the integration measure
\begin{eqnarray}
\widetilde{G}_{\alpha \beta}&=& {1 \over 2} \, \varepsilon_{\alpha
\beta \rho \tau }  \, G^{\rho \tau} \,, \qquad \int [{\cal D}
\alpha_i] \equiv \int_0^1 d \alpha_q \, \int_0^1 d \alpha_{\bar q}
\, \int_0^1 d \alpha_g \, \delta \left (1-\alpha_q - \alpha_{\bar
q} -\alpha_g \right )
\end{eqnarray}

%%%%%%%%%%%%%%%%%%%%%%%%%%%%%%%%%%%%%%%%%%%%%%%%%%%%%%%%%%%%%%%%%%%%%%%%%%%
%\newpage

\end{document}